\documentclass[12pt,letterpaper]{article}
\usepackage{amsfonts}
\usepackage{amssymb}
\usepackage{amsbsy}
\usepackage{latexsym}
\usepackage[dvips]{graphicx}
\evensidemargin=\oddsidemargin
\usepackage{epsfig}
\usepackage{citesort}

\newcommand{\beq}{\begin{equation}}
\newcommand{\eeq}{\end{equation}}
\newcommand{\bea}{\begin{eqnarray}}
\newcommand{\eea}{\end{eqnarray}}
\newcommand{\ba}{\begin{array}}
\newcommand{\ea}{\end{array}}

\newcommand{\nbk}{\bar{n}_k}

\newcommand{\bra}{\langle}
\newcommand{\ket}{\rangle}
\newcommand{\al}{\alpha}
\newcommand{\be}{\beta}
\newcommand{\e}{\epsilon}

\begin{document}
\begin{flushright}
hep-th/0306070\\
HIP-2003-36/TH \\
\end{flushright}
\vspace*{3mm}
\begin{center}
{\Large {\bf Holographic Bounds on the UV Cutoff Scale} \\[.1cm]
{\bf in Inflationary Cosmology}\\} \vspace*{12mm}
{\bf Esko Keski-Vakkuri}\footnote{\tt e-mail:
  esko.keski-vakkuri@helsinki.fi} {\bf and Martin S. Sloth}\footnote{\tt e-mail: martin.sloth@helsinki.fi}\\
\vspace{3mm}
{\it Helsinki Institute of Physics\\[.1cm]
P.O. Box 64, FIN-00014 University of Helsinki, Finland}
\vspace{1cm}
\begin{abstract}
\noindent We discuss how holographic bounds can be applied to the quantum
fluctuations of the inflaton. In general the holographic principle will
lead to a bound on the UV cutoff scale of the effective theory of
inflation, but it will depend on the coarse-graining prescription involved in
calculating the entropy. We propose that the entanglement entropy is a natural
measure of the entropy of the quantum perturbations, and show which
kind of bound on the cutoff it leads to. Such bounds are related
to whether the effects of new physics will show up in the CMB.
\end{abstract}
\end{center}
\setcounter{footnote}{0}
\baselineskip16pt

\section{Introduction and Summary}

Due to the increasing accuracy of cosmological data, there is
interest as to whether we can learn something about fundamental
physics from inflation
 \cite{Hassan:2002qk,Danielsson:2002kx,Bergstrom:2002yd,Kaloper:2002uj,Bozza:2003pr,transplanck}.
The holographic principle
is thought to be a generic feature of fundamental level
theories which include gravity, and might thus play an important model independent
role in such studies. The holographic principle
is basically a statement about the number of fundamental degrees of freedom.
On the other hand, it has lately become customary to view inflation as an
effective field theory, with a cutoff at energy scales where fundamental
physics becomes important. The number of degrees of freedom will depend on the
cutoff scale, hence holography might quite reasonably, in some way
or another, be connected with an upper bound on the cutoff scale of the
effective theory.

In this paper we explore this possibility in more detail. We
do not find any strong constraints from holography itself,
but it instead tells us something about the entropy of
cosmic perturbations. Only indirectly, does holography lead to a bound on
the cutoff scale. The bound appears to be weaker than the one in
\cite{Albrecht:2002xs}, and no stronger than one would expect
from ordinary back-reaction considerations. One issue that we shall
discuss is the
classical versus quantum nature of the entropy.
In previous studies of holography, the entropy of the cosmic perturbations
has usually been treated at classical level, based on some coarse-graining
prescription. However, the modes on sub-horizon scales are inherently quantum,
so any bounds on the entropy inside the horizon of a
cosmic observer should involve a semiclassical analysis\footnote{On an
even more sophisticated level, there are coarse-graining prescriptions taking
into account interaction with environment and the ensuing decoherence
\cite{Kiefer:1999sj}. For holography, one should then analyze bounds on 
the total entropy of cosmic perturbations plus the environment. However, 
previous studies of holography and the present study restrict to 
a toy model level, focusing only on the cosmic perturbations.}.
The holographic bounds are typically based on a statement about the
entropy of a matter system {\it before} it is lost across the horizon. It
then seems clear that it is the
entropy of the {\it quantum} modes, before they exit the horizon, that
should have a holographic bound. This being the case, we propose that
a natural concept of entropy of the modes on sub-horizon scales is the
entanglement entropy\footnote{Quantum
entanglement across the horizon in de Sitter
space has also been discussed in a different context in
\cite{Hawking:2000da}.}. It naturally scales like the horizon
area, and is thus often also called the geometric entropy. It also
depends explicitly on the cutoff scale.

We then move on to consider the shift of the horizon area, due to
the backreaction of the density perturbations on the geometry, or
due to the time evolution of the Hubble parameter during the slow
roll. In both cases, the change of the area leads to a flux of
entropy across the horizon. However, according to the holographic
bound, the entropy flux must be smaller than the slow-roll change
of the horizon area. This leads to an "indirect" bound on the
cutoff scale.

This paper is organized as follows. In sections 2 and 3 we first
review some basic facts of holographic bounds and the theory of
density perturbations. In the end of section 3 we discuss the
coarse-grained and the geometric entropy as the entropy of the
perturbations. In section 4 we analyze the backreaction of the perturbations
and the ensuing shift in the horizon area. In section 5 we discuss
the direct and the indirect entropy bound, and apply them to
density fluctuations in section 6. The different bounds are then
summarized in a figure in section 6.
We use the definitions $M_p=1/\sqrt{8\pi G_N}$,
$l_p=\sqrt{G_N}$, and set $\hbar =c=1$.

\section{Holography and Cosmological Density Perturbations}

If the entropy of a matter system could be arbitrary large while its
total energy would be fixed, the
generalized second law (GSL) could be violated.
As a {\it gedanken} experiment, drop the matter system
into a black hole. The
entropy of the matter system is lost, while the increase in the
entropy of the black hole is proportional to the increase of its horizon
area, which only depends on the total energy of the matter system.
Thus, if the entropy of the matter system was high enough, the
total entropy will decrease in the process. However, there are
strong indications
that the generalized second
law is a true principle of Nature. This led Bekenstein to suggest an
entropy bound on matter systems coupled to gravity.
It has later been developed into a more general entropy bound,
called the covariant entropy bound
\cite{Bousso:1999xy}.

One can apply the same reasoning in inflation where
cosmological perturbations carry entropy and redshift to superhorizon
scales. The {\it gedanken} experiment which led to
the entropy bounds on matter, suggests that one should consider the
dynamical process in which entropy is lost to the horizon and
compensated by an increase in the horizon area. This might lead to
interesting constraints in the theory of inflation.

The first natural application to consider is known as
the {\it D-bound}. Gibbons and Hawking have
shown that de Sitter horizon is associated with an entropy \cite{Gibbons:mu}
\beq
S_0=\frac{1}{4}A_0 = 4\pi\frac{1}{H^2}
 \eeq
where $A_0$ is the area of the de Sitter horizon and $H$ is the Hubble
parameter. Now, if we place a freely falling matter system
inside the horizon of an observer, she will see the matter system redshift
away. The horizon area grows, as the accessible space-time region
again converts into empty de Sitter space.
The entropy of the final state is $S_0$, while the entropy of the initial state is
the sum of the matter entropy $S_m$ and the Gibbons-Hawking
entropy, one quarter of the initial area $A_c$ of the horizon:
 \beq
S=S_m+\frac{1}{4}A_c~.
 \eeq
{}From the generalized second law ($S\leq S_0$), one obtains a bound
on the entropy of the matter
 \beq \label{D-bound}
S_m\leq(A_0-A_c)~.
 \eeq
This is called the D-bound \cite{Bousso:2000nf}.

The horizon area also changes during slow-roll inflation.
However, the change in the horizon area must be small,
for the slow-roll approximation to be valid. The expansion rate
of the horizon area is given by the slow-roll parameter $\e$. It is
natural to expect that the D-bound (\ref{D-bound}) will
lead to an $\e$-dependent upper bound on the entropy
carried by the cosmological perturbation
modes, as they redshift to superhorizon scales.
Whether this is a non-trivial bound, depends on how one calculates the
entropy of the modes. We will next review some of the
suggested methods, and suggest a new particularly natural one.

\section{Entropy of Perturbation Modes}

For convenience and to collect together some useful formulas,
we begin with a short review of the theory of density
perturbations.

\subsection{The Origin of Density Perturbations}

One can show that the evolution of scalar (or tensor)
perturbations is described by a minimally coupled massless scalar
field $\phi$. For brevity, we will only consider a spatially flat
FLRW universe, with the metric
 \beq
ds^2 =-dt^2+a^2(t)\delta_{ij}dx^idx^j~.
 \eeq
After the field redefinition $\mu\equiv a\phi$
the equation of motion for the Fourier modes $\mu_k$ in de Sitter space
takes the form
 \beq
\mu''_k+\left[\omega^2(k)-\frac{a''}{a}\right]\mu_k=0~,
 \eeq
where prime denotes derivative with respect to the conformal time $\eta$
defined as $dt\equiv a(t)d\eta$.

It is convenient to introduce a cutoff scale $\Lambda$ in Fourier space
to parameterize our ignorance of physics beyond this scale\footnote{Note
that the cutoff must be in physical momentum ($p<\Lambda$) in order
not to be macroscopic at the end of inflation. In comoving momentum the cutoff is
$k<a\Lambda$}. As a first approximation, one could consider it to
be of the order of the Planck scale $M_p$. However, the main issue is if
one can derive tighter bounds for it from holographic considerations.
The possible effect of such a cutoff on the observed CMB spectrum was
investigated in
\cite{Danielsson:2002kx,Danielsson:2003cn,Kaloper:2002uj,Bozza:2003pr}.
We will assume that the physics below the cutoff scale obeys the standard
linear dispersion relation $\omega^2(k)= k^2$.  Thus,  below the cutoff the
solution to the field equation is given by
 \beq
\mu_k(\eta)=\frac{\al_k}{\sqrt{2k}}e^{-ik\eta}+\frac{\be_k}{\sqrt{2k}}e^{ik\eta}~,
 \eeq
with $|\al_k|^2-|\be_k|^2=1$ from the usual Wronskian normalization
condition. The origin of density fluctuations is a quantum effect.
At quantum level the field $\mu$ is a pure state
\beq
    \tilde\rho = | 0, in\ket \bra 0, in | \ ,
\eeq
corresponding to initial absence of
density fluctuations. In the above case, where trans-Planckian
effects have been excluded (the dispersion relation is linear),
the initial vacuum $|0,in\ket$ is the one with respect to modes with
$\beta_k=0$, the Bunch-Davies vacuum.
However, due to the time-dependence of the background, at later times the
physical vacuum becomes a different one, with respect to $\mu_k$ above.
With respect to the new vacuum and particle states, the state $\tilde \rho$
appears as a many-particle (squeezed) state
\beq
  \tilde\rho = \prod_{k,k'}\sum^{\infty}_{n_k,m_{k'}=0}
        \tilde\rho (n_k,m_{k'})~|m_{k'}\ket \bra n_k | \ ,
\eeq
with the average occupation number per wavenumber
\beq
   \bar{n}_k = \sum^{\infty}_{n_k=0}
        \tilde\rho (n_k,n_k)~n_k
       = |\beta_k|^2 .
\eeq
These excitations correspond to density fluctuations with respect
to the new vacuum.
In de Sitter space, for a massless scalar field
in the initial vacuum, the average occupation number is found to
be
\beq \label{nbk}
 \nbk \sim (aH)^2/(2k)^2 \ .
\eeq

One can also introduce the squeezing parameter $r_k$
and denote
 \beq
\nbk =\sinh^2(r_k)~.
 \eeq
If $\nbk\gg 1$ (large squeezing), the population of modes is so large that they can
be approximated by a classical fluid. In the opposite limit, $\nbk \ll 1$,
the density perturbations behave quantum mechanically.

On sub-horizon scales the curvature $a'/a$ can be neglected
and the energy density is
 \beq \rho =
 \bra 0,in |T^0_{\ 0}|0, in\ket
=\frac{1}{4\pi^2a^4}\int_{aH}^{a\Lambda} dk
k^3\left(\frac{1}{2}+|\be_k|^2\right)~.
 \eeq
After subtracting the zero-point energy,
 \beq \label{rho}
 \rho = \frac{1}{4\pi^2 a^4}\int_{aH}^{a\Lambda} dk
k^2\omega(k) \bar{n}_k~.
\eeq
Substituting the result (\ref{nbk}) in (\ref{rho}),
one finds
\beq \label{re0}
\rho
=\frac{1}{32\pi^2}\Lambda^2 H^2~.
\eeq

Inside the horizon, when $|k\eta|\gg 1$ and the space-time is
effectively flat, the equation of state becomes
$p=(1/3)\rho$, that of a relativistic fluid or a free massless
scalar field in Minkowski space.
Since the density fluctuations now behave like a fluid, it becomes
natural to associate an entropy with the mean
occupation number.
However, strictly speaking the ensemble is still in a pure state
with zero entropy, so the entropy of the fluid
will involve some amount of coarse graining.

\subsection{The Entropy of the Density Perturbations}

The aim of this paper is to apply holographic entropy bounds to the
inflationary quantum fluctuations. 
However, it is clear that the modes
are not classical before they exit the horizon and one
should be careful which entropy is assigned to those modes. The
entropy bounds, like the D-bound, are bounds on the true degrees of
freedom of the fundamental theory. When they are applied to the
inflationary quantum fluctuations,
unnecessarily strong constraints may result simply due to excessive
coarse-graining.
We now take a brief look at the standard coarse-graining
prescriptions.

\paragraph{Coarse-grained entropy.}
A popular approximation in the literature is that on superhorizon scales,
where the average occupation number is much greater than one, the entropy per
comoving momentum is given in terms of the squeezing parameter $r_k$ as
 \beq \label{Sk}
S_k=2r_k~,
 \eeq
since it agrees with the classical concepts
of entropy. In \cite{Brandenberger:1992jh}, it
was proposed that the entropy of a quantized field with average occupation
number $\nbk$ is in general given by
 \beq \label{Br}
S=\sum_k g_k\left[(\nbk +1)\ln(\nbk +1)-\nbk \ln \nbk \right]~,
 \eeq
where $g_k$ is the degeneracy. This expression for the entropy
has the same form as that of a thermal distribution, and it
reproduces equation (\ref{Sk}) in the limit of large squeezing
$r_k \gg 1$.  It is tempting to take the limit
$\nbk \ll 1$. With $g_k=(4\pi/3a^3)Vk^2dk$, one finds
 \beq
S=-\frac{4\pi V}{a^3}\int d^3k~\nbk \ln \nbk~.
 \eeq
or, alternatively, the entropy per mode $S_k\sim -\nbk \ln \nbk$.
However, the considerations in \cite{Brandenberger:1992jh} do not
necessary apply in the $r_k\ll 1$ limit.
References \cite{Gasperini:1993mq,Gasperini:1992xv} followed
another coarse graining approach.
It was argued that the expression for
entropy per mode in equation (\ref{Sk}) is valid for all values of the
squeezing parameter, $r_k$.

\paragraph{Entanglement entropy.}
The reference \cite{Danielsson:2003cn} looked for a more
detailed understanding by going back to the origin of the entropy.
It was argued that from the point of a local
observer there is a natural bound on the freedom to coarse-grain
the system. Imagine that the full system is divided into two
subsystems, with $N_1$ and $N_2$ degrees of freedom, such that
the first subsystem and degrees of freedom are accessible to the local
observer. If the total system is in a pure state, when we trace
over the second subsystem (inaccessible to the local observer),
we find that the entropy of the second subsystem is bounded by the number of
the traced-over degrees of freedom,
$S_1<\ln N_2$. So it was concluded that from the local point of
view the production of entropy is limited by the ability to coarse
grain.

Let us elaborate this viewpoint.
What is essentially proposed in the above argument is that the maximum
entropy of the observable subsystem (the density perturbations
within the horizon) is defined as the entropy of entanglement, or the
geometric entropy \cite{geoment}. Trace over the
degrees of freedom outside the horizon
radius $r_H$ and define a reduced density
matrix
\beq
  \tilde\rho_{in} = Tr_{r>r_H} (\tilde\rho) ,
\eeq
this only describes the system within the horizon. But now $\tilde\rho_{in}$
is a mixed state, with an associated entropy
\beq
  S_{in} = -Tr_{r<r_H} (\tilde\rho_{in}\ln \tilde\rho_{in}) \ .
\eeq
M\"uller and Lousto have shown \cite{Muller:1995mz} that the entropy is
\beq
\label{Sgeom}
  S_{in} \approx 0.3~(r_H /\varepsilon)^2
\eeq
where $r_H=1/Ha$ is the radius in the comoving frame and $\varepsilon$
is an UV cutoff scale defining the sharpness of the 'cut'. The
cutoff scale is of course ambiguous, making the entropy
ill-defined and formally divergent as the scale approaches zero.
However, in the present context, in our treatment of the
density fluctuations, we have already explicitly introduced an
UV cutoff $\Lambda$ in the Fourier space, and the entropy
or the fluctuations is expected to have an
explicit dependence of that scale. Therefore it is logical to
set\footnote{It would also be interesting to perform a refined
derivation of the entanglement entropy formula in the context where
trans-Planckian modifications have been incorporated into field
theory, for a more sophisticated analysis of the cutoff
dependence. Indeed, such a consideration has just appeared
\cite{CCL}.}
$\varepsilon =1/a\Lambda$. (The cutoff is imposed in the
physical frame, so in the comoving frame a scale factor is
included.)

One virtue of the entanglement entropy is that $S_{in}$ is
now directly associated with the degrees of freedom inside the horizon.
The "flow" of entropy across the horizon can now be
viewed as follows.
Consider a set of perturbations with comoving wavenumber $k>k_0$.
At very late times, they have essentially
all crossed well outside the horizon, see Figure 1, and can then be
approximated by a classical fluid. On the other hand, at very early times
($a<k_0/H$) they are all within the horizon of (comoving) radius
$r_H = 1/Ha$. Then the average occupation number per mode is of the
order of one, and the system is far from classical. At this point
it is natural to describe the system by $\tilde\rho_{in}$ as above, so
the initial entropy assigned to these seeds of perturbations
inside the horizon is (\ref{Sgeom}).
Consider then the same set of density perturbations at
late times, when they are (practically) all grown
outside the horizon (Figure 1). Now, in order to describe the degrees of
freedom outside the horizon, it is natural to
trace over the
interior of the horizon to obtain a reduced density matrix $\tilde\rho_{out}$
for the outside degrees of freedom. The resulting entropy turns
out to be the same:
\beq
 S_{out}=S_{in}  \equiv S_{ent} ~.
\eeq
Naively the entropy appears
to have decreased, since we have been using the comoving frame
where $r_H=1/Ha$ decreases in time. However, in the physical frame
the entropy is a constant,
\beq
 S_{ent} = 0.3~\left( \frac{ \Lambda }{ H} \right)^2 \ .
\eeq

\begin{figure}[hbtp] \label{ent}
\begin{center}
\includegraphics[height=7cm]{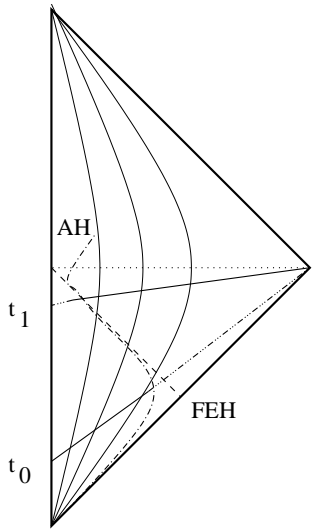}
\end{center}
\caption{\small In the Penrose diagram above, modes corresponding to constant
  comoving radius which are inside the apparent horizon (AH) at time $t_0$ are
  shown. (The future event horizon (FEH) is also depicted in the
  figure.) The diagram shows the inflation, radiation and matter
  dominated eras. It is illustrated that calculating the entropy of those modes
  by tracing over the exterior of the apparent horizon on the time slice
  $t_0$, is equivalent to later calculating their entropy by tracing over
  the interior of the horizon at time $t_1$.}
\end{figure}

As the perturbations have grown well outside, the average
occupation number per mode $\nbk \gg 1$ and
they can be approximated by a classical fluid. We could then
assign the above entropy for the fluid. But, at earlier times
when the modes were inside the horizon, they had the same entropy.
Let us compare $S_{ent}$ with the coarse-grained entropy in the
sub-horizon region following the prescription of
Giovaninni and Gasperini \cite{Gasperini:1992xv} (See also
\cite{Kruczenski:1994pu}). They suggested that the
entropy per comoving mode is given by $S_k=2r_k$. For small
squeezing, $r_k\sim|\be_k|=aH/(2k)$, hence the entropy contained in the
relativistic particles sub-horizon ($n_k \ll 1$) is given by
\beq
S_{cg}=\frac{V}{\pi^2a^3}\int_{aH}^{a\Lambda}dk~k^2~\frac{aH}{2k}
\simeq \frac{1}{3\pi} \left( \frac{\Lambda }{H} \right)^2~,
\eeq
where $V=(4/3)\pi H^{-3}$. This is in agreement with the
entanglement entropy, up to an irrelevant numerical factor of order
one. In the coarse-grained approach it is not evident why the
prescription should be valid in the sub-horizon region where the
population of the modes is small (and the system is essentially
quantum). Further, the coarse-graining process drops out
information of the system which is not manifestly associated with
a region of spacetime. However, the fact that the result agrees
with the entanglement entropy gives it further credibility. The
coarse-grained entropy saturates the bound arising from the
entanglement of the subsystems\footnote{On the other hand, in a more
realistic analysis where one takes into account other fields as an 
environment, the 
decoherence-induced coarse-grained entropy of the cosmic perturbations 
is less \cite{Kiefer:1999sj} than the bound. But then a proper
count of the total entropy should include the contribution from the environment
as well.}.

To summarize, we considered density perturbations at early and late
times. At late times they are essentially classical, outside
the horizon, and it is natural to
describe their entropy as that of a classical fluid (defined by a suitable
coarse graining). However, we argued that considering the quantum mechanical
origin of the perturbations, the most natural notion of entropy is actually
the entanglement entropy -- and that one is the same as the entropy
of the perturbations at very early times when they are all within the
horizon.
We just choose to trace
over opposite regions of space.
The various definitions of coarse-grained entropy
must be consistent with the entanglement entropy,
which in turn naturally scales like the area. Then, it is natural
that the holographic bound is satisfied. In the prescription for
the entropy, it is also explicit why there is a bound on the
number of degrees of freedom inside the horizon. This in turn is
somewhat less satisfying in the coarse-graining descriptions: first
one performs a coarse-graining procedure which makes no reference
to a specific region in spacetime, then the "put-in-by-hand" entropy should
correspond to at most a specific number of degrees of freedom inside
the horizon dictated by the holographic bound.

\section{Perfect Fluid with a Cutoff in Quasi de Sitter Space}

In the previous section, we showed that the entropy of the
inflationary quantum fluctuations scales like the horizon area times
the square of the UV cutoff.
Next, we look at the backreaction and the shift in horizon area
due to the loss of fluctuations across the horizon. Before
introducing the entropy bounds in section 4, it is useful to discuss
this relationship in more detail.

To this end, let us review some of the discussion in
\cite{Frolov:2002va} (see also \cite{Jacobson:1995ab}). It is argued
that the change in the horizon area is given by the energy flux
through the horizon $\delta E$ according to the first law of
thermodynamics
 \beq
TdS = \delta E~,
 \eeq
where $S$ is the geometrical entropy.
In \cite{Frolov:2002va} the energy flux is written as
 \beq
\delta E = \delta \int d\Sigma_{\mu}T_{\nu}^{\mu}\zeta^{\mu}~,
 \eeq
where $d\Sigma_{\mu}$ is the 3-volume of the horizon and in the
metric
\beq
ds^2=-dt^2+e^{2Ht}\left(dr^2+r^2d\Omega^2\right)~,
\eeq
$\zeta^{\mu}=(1,-Hr,0,0)$ is the approximate Killing vector which is
null at the horizon. The coordinate position of the horizon is given
by $r=1/aH$. By equating $\delta E/T$ to the change in the
geometrical entropy given by the variation
in the horizon area, one obtains
 \beq
\dot{H}=-4\pi G T_{\mu\nu}\zeta^{\mu}\zeta^{\nu}~.
 \eeq
It was then shown that for the energy momentum tensor of the background inflaton
field, the energy flux through the horizon is
$T_{\mu\nu}\zeta^{\mu}\zeta^{\nu}=\dot{\phi}$, which leads to the
usual Einstein equation
 \beq
\dot{H} = -4\pi G\dot{\phi}^2~.
 \eeq
On other hand, the density fluctuations can be viewed as a perfect fluid with
the energy momentum tensor
 \beq
T_{\mu}^{\nu} = diag(-\rho,p,p,p)~,
 \eeq
with $p=\rho /3$. In this way one finds
 \beq
T_{\mu\nu}\zeta^{\mu}\zeta^{\nu}=\rho+p~.
 \eeq
As expected, a cosmological constant $\rho=-p$ leads to no change in
the horizon area, while a constant energy density $p=(1/3)\rho$ leads to
 \beq \label{m1}
\dot H = -\frac{16\pi}{3} G \rho~.
 \eeq
This agrees with what we would expect from the Einstein equations.
Introducing a small number $\tilde\epsilon$,
we can parameterize the energy density as
 \beq \label{re}
\rho = \frac{\tilde \epsilon}{2}3H^2M_p^2~.
 \eeq
Then, (\ref{m1}) can be rewritten in the form
 \beq \label{qwq1}
\dot H \simeq -\tilde\e H^2~.
 \eeq
The energy flux across the horizon from the constant
radiation fluid leads to a small slow-roll-like change in the
horizon area.

There might seem to be a contradiction
between the continuity equation
 \beq
\dot{\rho}+3H(\rho+p)=0~,
 \eeq
which implies that a
constant energy density is always associated by a negative but
equal pressure, and the equation of state of a massless scalar field
$p=(1/3)\rho$. However, one should note that the energy
conservation equation is violated by the fixed cutoff,
since the cutoff acts as an energy reservoir from
which energy is pumped into the system.

This can also be understood in terms of a source term. For a constant energy density,
${T^{\mu0}}_{;\mu}=0$ would imply
that $\rho=-p$ and $\dot H=0$. But obviously ${T^{\mu\nu}}_{;\mu}=0$ is not valid since
modes redshift across the cutoff and energy in this way leaks into the
system that is no longer closed. In fact, we should include a source
term modifying the continuity equation to
${T^{\mu0}}_{;\mu}=-4H\rho$.

Heuristically, even if the energy density is constant, the energy
density of relativistic
particles which were present at any given time $t'$ will fall off
like $1/a^4$ and thus behave like relativistic fluid. Those modes
will eventually redshift away and if no new modes had been
created to sustain the pressure, the horizon area would have
changed by $\Delta A$ which then limits their entropy. Even if new
modes are created continuously to make up a constant energy density,
we have seen that it is natural to assume that the modes
that do redshift (exit the horizon), will change the horizon area
proportional to the
entropy they carry according to the generalized second law\footnote{A
related application of the GSL was considered also in
\cite{Brustein:1999ua}.}. This change must be smaller than the
observed slow-roll change. This is the indirect bound that we will
discuss in the next section.

\section{Entropy Bounds}

Now we are ready to discuss two specific entropy bounds on
inflationary quantum fluctuations; a {\it direct} and an {\it indirect}
entropy bound. The first one is just an application of the D-bound
introduced in section 2, and as we will argue, it could be viewed
as a consistency check which is trivially satisfied if one has performed
the right amount of coarse graining. The second one is a bound on
the inflationary quantum fluctuations from the known background
evolution of the geometry.

\subsection{The Direct Entropy Bound}

Consider a massless scalar field with
the energy density inside the horizon (\ref{rho}).
The energy density carried in these modes must be much smaller
than the critical energy density. Like in equation (\ref{re}) above,
it is useful to parameterize the energy density in terms of
$\tilde\e \ll 1$.

In \cite{Danielsson:2003cn} it was shown that this amount of relativistic
matter embedded in a pure de Sitter space would change the
horizon by $\Delta A$ where
 \beq
\Delta A=A_f-A_0=\frac{\tilde\epsilon}{2}A_f~.
 \eeq
Here $A_0$ is the horizon area of the initial de Sitter space with
$\rho$ included and $A_f=4\pi H^{-2}$ is the horizon area of the
final pure de Sitter space after the matter component as exited
the horizon. The D-bound \cite{Bousso:2000md} then implies that the
entropy contained in the matter is limited by
 \beq \label{DA}
S\leq \frac{1}{4}\frac{\Delta
A}{l_p^2}=4\pi^2\tilde\e\frac{M_p^2}{H^2}~.
 \eeq

By using equation (\ref{re}), we can
eliminate $\tilde \e$. The D-bound (direct bound) then takes the
form
 \beq \label{dib}
S\leq \frac{8\pi^2}{3}\frac{\rho}{H^4}
 \eeq
Using that the vacuum energy of the de Sitter space is
$\rho_{\Lambda}=3H^2 M_p^2$, we can write the direct
bound in equation (\ref{dib}) in a more transparent way
 \beq
S\leq 24\pi^2
\frac{M_p^4}{\rho_{\Lambda}}\frac{\rho}{\rho_{\Lambda}}~.
 \eeq

Now, the D-bound should not be viewed as something intrinsically related to de Sitter
space, but only a way to use de Sitter space to obtain a bound on the
entropy of matter, much like the Schwarzschild solution leads to the
Bekenstein bound. Thus, a reasonable choice of coarse graining should
render the bound in equation (\ref{dib}) trivial. If one obtains a
nontrivial constraint from the bound in equation (\ref{dib}), it is a hint
that the counting of entropy is unphysical. Indeed, the geometrical
entropy of the inflationary quantum fluctuations trivially satisfies
the bound. We will return to the examples in section 6.

\subsection{The Indirect Entropy Bound}

In a slow-roll set-up, inflation is driven by the inflaton
potential $V$, and
 \beq
H^{-1}=\frac{M_p}{\sqrt{3V}}~.
 \eeq
During slow-roll the potential is almost constant and in one
e-fold, it changes maximally by
 \beq
\Delta V\simeq \frac{\epsilon}{2}V~,
 \eeq
where $\e=1/2~(V'/V)^2$ is the slow-roll parameter and prime
denotes derivative with respect to the inflaton field $\phi$.
This also implies that during one e-fold the horizon area changes
by
 \beq
\Delta A \simeq \frac{\epsilon}{2}A=4\pi \frac{\e}{H^2}~.
 \eeq
The entropy that exits the horizon during one e-fold\footnote{The
  entropy that exits the horizon is equal to the loss of entropy
  inside the horizon due to redshift. See the discussion after
  eq.(\ref{s9}) in the appendix.} is then
limited by
 \beq \label{Aind}
\Delta S_{exit}\lesssim \frac{1}{4}\frac{\Delta
A}{l_p^2}=8\pi^2\e\frac{M_p^2}{H^2}~.
 \eeq
This is what we refer to as the indirect entropy bound. It is of
course weaker than the D-bound, since it implicitly assumes $\tilde
\e\ll \e$. However, it bounds the flux of entropy across the horizon
by the actual change in the geometry in a cosmological setting. This
analogous to some of the considerations in also \cite{Albrecht:2002xs}
and \cite{Brustein:1999ua}.

\section{Applications to Quantum Fluctuations in the Early Universe}

In this section we apply the considerations above explicitly to the
theory of cosmological perturbations in quasi de Sitter space.

As a first very trivial consistency check, consider the coarse-grained
or entanglement entropy
 \beq
S \simeq C\cdot \left( \frac{\Lambda}{H}\right)^2 ~,
 \eeq
where $C$ is a numerical factor of order
one, and the D-bound. It can be seen by comparing equations (\ref{re0}) and
(\ref{dib}), that the D-bound is trivially satisfied if $C$ is of the
order one. The only difference between the bound applied to the
geometric versus
to the coarse-grained entropy is a conceptual one. In the latter
case the bound is a more arbitrary consistency check. There seemed
to be no direct understanding as to why the discarded information
in the coarse-graining scheme should be related to a holographic
bound on the degrees of freedom inside the horizon. In the case of
the geometric entropy, the correspondence is more transparent.

The indirect bound is a bit more interesting. From equations
(\ref{s7}) and (\ref{s8})
it is easy to see that in one e-folding, the amount of
entropy that exit the horizon is given by
 \beq
\Delta S_{exit} =\frac{1}{\pi}\frac{\Lambda^2}{H^2}~.
 \eeq
{}From the indirect entropy bound in equation (\ref{Aind}), we obtain\footnote{Note that if the total entropy that has left the horizon during the
first few e-foldings is given by $S\simeq 4\pi\Lambda^3/H^3$ and if it
dominates the total entropy in modes that have
left the horizon, as argued in \cite{Albrecht:2002xs}, then we reproduce the
constraint in the cutoff in \cite{Albrecht:2002xs}
$\Lambda\lesssim \e^{1/3}\left(HM_p^2\right)^{1/3}$.
This estimate was obtained by using, as a measure for the entropy in
the lost modes, their entropy just as they stretch to superhorizon
scales. Note, however, that in the coarse-graining approach the loss
of entropy due to redshift is distributed on all
modes inside the horizon, as also discussed in the appendix.}
 \beq
\Lambda\lesssim  \sqrt{8\pi^3\e}M_p
 \eeq
Using equation (\ref{re0}), one can see that the indirect bound in
this case turns out to be similar to requiring $\tilde\e\lesssim \e$,
with the definition of $\tilde\e$ given in equation (\ref{re}). Note
that this result depends on the
coarse-graining prescription. If one approximate the entropy per mode
by an almost constant, one reproduces the stronger bound of Albrecht
and Kaloper \cite{Albrecht:2002xs} (see also the footnote).

This is an interesting bound, but in many cases $\e\sim 1/N$ where $N$ is
the number of e-foldings till the end of inflation. If inflation lasts
about 100 e-foldings, this bound is no stronger than $\Lambda \ll M_p$.
On the other hand, if inflation lasts much longer than 100
e-foldings then the bound is potentially interesting. The bound is
illustrated in figure 2.

\begin{figure}[thbp] \label{holo}
\begin{center}
\includegraphics[width=7cm]{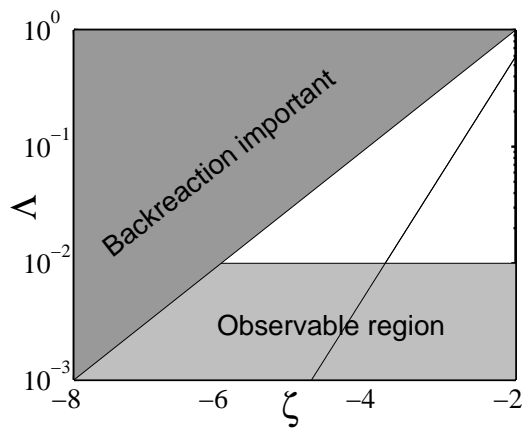}
\end{center}
\caption{\small The slow-roll parameter $\e=10^{\zeta}$ vs. the cutoff scale
  $\Lambda$. For $H\sim 10^{-4}M_p$, the range in which
  trans-Planckian effects are observable are also indicated. It is
  argued in \cite{Bergstrom:2002yd} that if $H/\Lambda>10^{-2}$, then
  the trans-Planckian effects can be observable. If the
  cutoff scale is too high, the perturbations become important for the
  background evolution. This is indicated by the dark-gray region. The
  region below the additional line is the region allowed by the bound
  obtained from the $\al$-vacua scenario.}
\end{figure}

We also note that, if the vacuum is chosen to be the Minkowski vacuum
for each mode as they exits the trans-Planckian regime as in \cite{Danielsson:2002kx},
such that
 \beq
|\be_k|\simeq \frac{1}{2}\frac{H}{\Lambda}
 \eeq
on sub-horizon scales, then the total entropy inside the horizon,
calculated by using $S_k=2r_k$, will still scale as
$\Lambda^2/H^2$. Also the energy density will still be given by
equation (\ref{re0}). Hence, the direct and the indirect bounds will
not change significantly.

As a final example we mention the $\alpha$-vacuum as suggested in
\cite{Goldstein:2002fc} (for a discussion of the $\alpha$-vacuum, see
\cite{alphavac}). It was suggested that the modes below the cutoff should
be placed in the $\alpha$-vacuum with a trans-Planckian signature
determined by
 \beq
|\be_k|\simeq \frac{HM_p}{\Lambda^2}~.
 \eeq
In the present set-up, the direct bound would again be trivial, while
the indirect bound leads to the more interesting constraint
 \beq
 \Lambda \lesssim 6\pi^2\e M_p~.
 \eeq
This constraint is also illustrated in figure 2.

In the present approach we have ignored the modes above the
cutoff. The above-the-cutoff region
instead acts as a reservoir. If one has an explicit model for incorporating
the trans-Planckian physics, like in the
approach of \cite{Hassan:2002qk}
or \cite{transplanck}, one can in principle avoid the use of the
explicit cutoff.
Instead it will appear as the scale controlling the strength of the
modifications of the standard quantum field theory.

For example in \cite{Hassan:2002qk}, the effect of fundamental physics
on the evolution of perturbation modes
above the cutoff was modeled by incorporating effectively a
minimum length (mimicking string theory) into to the field
theory at the level of modified commutation relations. Using this
approach, it was also discussed how the modes are
created in the trans-Planckian regime as an effect of the novel feature of
trans-Planckian damping.

Since the modes above the new physics scale will
also contribute to the energy and entropy inside the horizon, we expect
that the indirect bound on the new physics scale will be at least
marginally more tight depending on the model. This is a possible
direction for further investigation.

\vspace{.5cm}
\noindent{\Large{\bf Acknowledgment}}
\vspace{.3cm}
\par\noindent
We would like to thank Kari Enqvist for suggesting our figure 2, and
Riccardo Sturani and Shinsuke Kawai for many helpful discussions. We
would also like to thank Raphael Bousso for helpful comments on the
first version, and A. Starobinsky for some illuminating
remarks. E.K-V. has been in part supported by the Academy of Finland. 
\appendix

\section{Relation to covariant entropy bound}

Since the covariant entropy bound is supposed to be the generalization
of the entropy bound discussed until now, it is interesting to briefly
exemplify how it is related to the discussion above.

Following \cite{Flanagan:1999jp,Bousso:2003kb,Strominger:2003br}, the
integral of the entropy flux $s$ over the lightsheet between two
spatial surfaces $B$ and $B'$ can be written as
 \beq \label{s1}
\int_{L(B-B')}s= \int_Bd^2x\sqrt{h(x)}\int_0^1d\lambda
s(x,\lambda)\mathcal{A}(x,\lambda)~.
 \eeq
We have chosen a coordinate system $(x^1,x^2)$ on $B$ and $h(x)$ is
the determinant of the induced metric on $B$. The affine parameter
$\lambda$ is normalized such that it takes the value one at $B'$. The
function $\mathcal{A}(x,\lambda)$ is the area decrease factor for the
geodesic that begins at the point $x$ on $B$.

As explained also in \cite{Flanagan:1999jp,Bousso:2003kb,Strominger:2003br},
the physical meaning is simply as follows: As we parallel propagate a
infinitesimal area $d^2x\sqrt{h(x)}$ from the point $(x,0)$ on $B$ to
the point $(x,\lambda)$ on the lightsheet, the area contracts to
$d^2x\sqrt{h(x)}\mathcal(x,\lambda)$. Thus the proper infinitesimal
volume on the lightsheet is $d\lambda
d^2x\sqrt{h(x)}\mathcal(x,\lambda)$. The infinitesimal volume times the
entropy flux, all of it integrated, then gives the entropy over the
lightsheet. As stated in
\cite{Flanagan:1999jp,Bousso:2003kb,Strominger:2003br}, the 
generalized covariant bound now takes the form
 \beq \label{s2}
\int_0^1d\lambda
s(\lambda)\mathcal{A}(\lambda)\leq\frac{1}{4}(1-\mathcal{A}(1))
 \eeq
for each geodesic of the lightsheet.

Now let us make that into a differential bound and define
 \beq
A_B =\int_Bd^2x\sqrt{h(x)}~.
 \eeq
Now assume that $\mathcal{A}(\lambda)$ and $s(\lambda)$ does not
 depend on $x$. Then we can write equation (\ref{s1}) as
 \beq \label{s3}
\int_{L(B-B')}s= A_B\int_0^1d\lambda
s(\lambda)\mathcal{A}(\lambda)~.
 \eeq
and equation (\ref{s2}) can be written
 \beq \label{s4}
A_B\int_0^1d\lambda
s(\lambda)\mathcal{A}(\lambda)\leq\frac{1}{4}A_B(1-\mathcal{A}(1))
 \eeq
or in infinitesimal form
  \beq \label{s5}
A_B s(0)\mathcal{A}(0)d\lambda \leq \frac{1}{4}dA_B~.
 \eeq

Since the matter is at rest, we can construct the entropy flux vector
$s_a$ by multiplying $\sigma$ with the four-vector $u_a=(-1,0,0,0)$,
 \beq
s_a=\sigma u_a.
 \eeq
Analogous to how we calculated the energy flux across the horizon, we
calculate the entropy flux across the horizon by multiplying $s_a$ by
the tangent vector to the null congruence generating the horizon
$k^a$. It is related to the approximate Killing vector
$\zeta^a=-H\lambda k^a$, where $\zeta^a=(1,-Hr,0,0)$ in
spherical coordinates. Thus, the flux across the horizon is $s = s_a k^a=
\sigma/(H\lambda)$. Using $d\lambda\propto adt$, as can easily be
verified, one can apply $d\lambda/\lambda=H dt$ in order to obtain
 \beq \label{s6}
A_B \sigma  \leq\frac{1}{4}\frac{dA_B}{dt}~.
 \eeq

Now let us apply the equation above to our cosmological setup. Let the
total entropy inside the horizon be denoted by $S$. Then we can
construct the entropy density $\sigma$ by dividing by the Hubble
volume
 \beq
\sigma=\frac{3H^3 S}{4\pi}~.
 \eeq
Inserted into equation (\ref{s6}), we find using $A_H=4\pi H^{-2}$,
 \beq \label{s7}
3HS \leq\frac{1}{4}\frac{dA_H}{dt}~.
 \eeq
The left-hand-side is basically the total entropy flux across the horizon,
so equation (\ref{s7}) is analogous to the indirect bound in equation
(\ref{Aind}) if the dominant time-dependence of the horizon area is
taken to be due to the slow-roll variation.

To compare, we consider how entropy is red-shifted away inside the
horizon. If one considers the entropy present in the radiation fluid
at some given moment, that entropy would dilute like $\sim a^{-3}$ if
no new entropy was generated. Thus the entropy density would redshift
as $a^{-3}$. This agrees with the familiar
statement that the entropy density of an equilibrium gas in an
expanding universe, dilutes as $a^{-3}$ as can be seen from
conservation of entropy per comoving volume $S_c$ in an adiabatic evolution.

This means that if the total entropy inside the horizon is given by
$S$, then there will be a loss in entropy inside the horizon due to
the red-shifting
 \beq \label{s8}
-\frac{dS}{dt}=3HS
 \eeq
Now comparing equation (\ref{s7}) and equation (\ref{s8}) we obtain
 \beq \label{s9}
-\frac{dS}{dt}\leq\frac{1}{4}\frac{dA_H}{dt}
 \eeq
which is just the generalized second law. Thus, for our particular
case, the generalized covariant bound reduces to the GSL which
underlies all of our discussion. In particular we see that the flow of
entropy through the horizon is the same as the loss of entropy due to
redshift inside the horizon (Compare left-hand-side of equation
(\ref{s7}) with the right-hand-side of equation (\ref{s8})).

Similarly we can, as a consistency check, show that the redshift of
modes inside the horizon agrees with the energy flux through the
horizon area. The energy-density of radiation redshift as $a^{-4}$, such
that the loss in energy density due to the red-shifting is
 \beq
-\frac{d\rho}{dt}=4H\rho
 \eeq
Thus, the total loss in energy due to red-shift inside the horizon is
 \beq \label{r1}
\frac{dE}{dt}=V\frac{d\rho}{dt} =\frac{-16\pi}{3}H^{-2}\rho~.
 \eeq
On the other hand, we calculated in section , that the energy flux
across the horizon is
 \beq
T_{\mu\nu}\zeta^{\mu}\zeta^{\nu}=\rho+p=\frac{4}{3}\rho~,
 \eeq
where we used the equation of state for radiation. This means that the
total flux of energy across the horizon is given by
 \beq \label{r2}
\frac{dE}{dt}=A_H T_{\mu\nu}\zeta^{\mu}\zeta^{\nu}= (4\pi
H^{-2})\frac{4}{3}\rho
=\frac{16\pi}{3}H^{-2}\rho
 \eeq
Again the expression in equations (\ref{r1}) and (\ref{r2}) agree.


\end{document}